\AtBeginDocument{%
  \providecommand\BibTeX{{%
    \normalfont B\kern-0.5em{\scshape i\kern-0.25em b}\kern-0.8em\TeX}}}

\documentclass[sigconf]{acmart}
\usepackage{subfigure}
\usepackage{subscript}
\usepackage{soul}

\copyrightyear{2020} 
\acmYear{2020} 
\setcopyright{acmcopyright}\acmConference[KDD '20]{Proceedings of the 26th ACM SIGKDD Conference on Knowledge Discovery and Data Mining}{August 23--27, 2020}{Virtual Event, CA, USA}
\acmBooktitle{Proceedings of the 26th ACM SIGKDD Conference on Knowledge Discovery and Data Mining (KDD '20), August 23--27, 2020, Virtual Event, CA, USA}
\acmPrice{15.00}
\acmDOI{10.1145/3394486.3403343}
\acmISBN{978-1-4503-7998-4/20/08}

\usepackage{multirow}
\usepackage{arydshln}
\usepackage{float}
\newcommand{\nop}[1]{}
\usepackage[symbol]{footmisc}

\begin{document}
\fancyhead{}

\title{Mining Implicit Relevance Feedback from User Behavior for Web Question Answering}


\author{Linjun Shou}
\authornotemark[2] 
\email{lisho@microsoft.com}
\affiliation{%
  \institution{STCA NLP Group, Microsoft, Beijing}
}

\author{Shining Bo}
\email{shiningbo@bupt.edu.cn}
\affiliation{%
  \institution{School of Information and Communication Engineering, BUPT}
}

\author{Feixiang Cheng}
\email{fecheng@microsoft.com}
\affiliation{%
  \institution{STCA NLP Group, Microsoft, Beijing}
}

\author{Ming Gong}
\email{migon@microsoft.com}
\affiliation{%
  \institution{STCA NLP Group, Microsoft, Beijing}
}

\author{Jian Pei}
\authornotemark[3]
\email{jpei@cs.sfu.ca}
\affiliation{%
  \institution{School of Computing Science, Simon Fraser University}
}

\author{Daxin Jiang}
\email{djiang@microsoft.com}
\affiliation{%
  \institution{STCA NLP Group, Microsoft, Beijing}
}

\renewcommand{\shortauthors}{Trovato and Tobin,~\textit{et al.}}

\begin{abstract}

\footnotetext[2]{Correspondence author.}
\footnotetext[3]{Jian Pei's research is supported in part by the NSERC Discovery Grant program. All opinions, findings, conclusions and recommendations in this paper are those of the authors and do not necessarily reflect the views of the funding agencies.}

Training and refreshing a web-scale Question Answering (QA) system for a multi-lingual commercial search engine often requires a huge amount of training examples. One principled idea is to mine implicit relevance feedback from user behavior recorded in search engine logs. All previous works on mining implicit relevance feedback target at relevance of web documents rather than passages. Due to several unique characteristics of QA tasks, the existing user behavior models for web documents cannot be applied to infer passage relevance. In this paper, we make the first study to explore the correlation between user behavior and passage relevance, and propose a novel approach for mining training data for Web QA. We conduct extensive experiments on four test datasets and the results show our approach significantly improves the accuracy of passage ranking without extra human labeled data. In practice, this work has proved effective to substantially reduce the human labeling cost for the QA service in a global commercial search engine, especially for languages with low resources. Our techniques have been deployed in multi-language services.


\end{abstract}



\maketitle

\section{Introduction}

\begin{figure}[t]
\setlength{\belowcaptionskip}{-0.2cm}
\centering
\includegraphics[scale=0.54, viewport=100 270 600 470, clip=true]{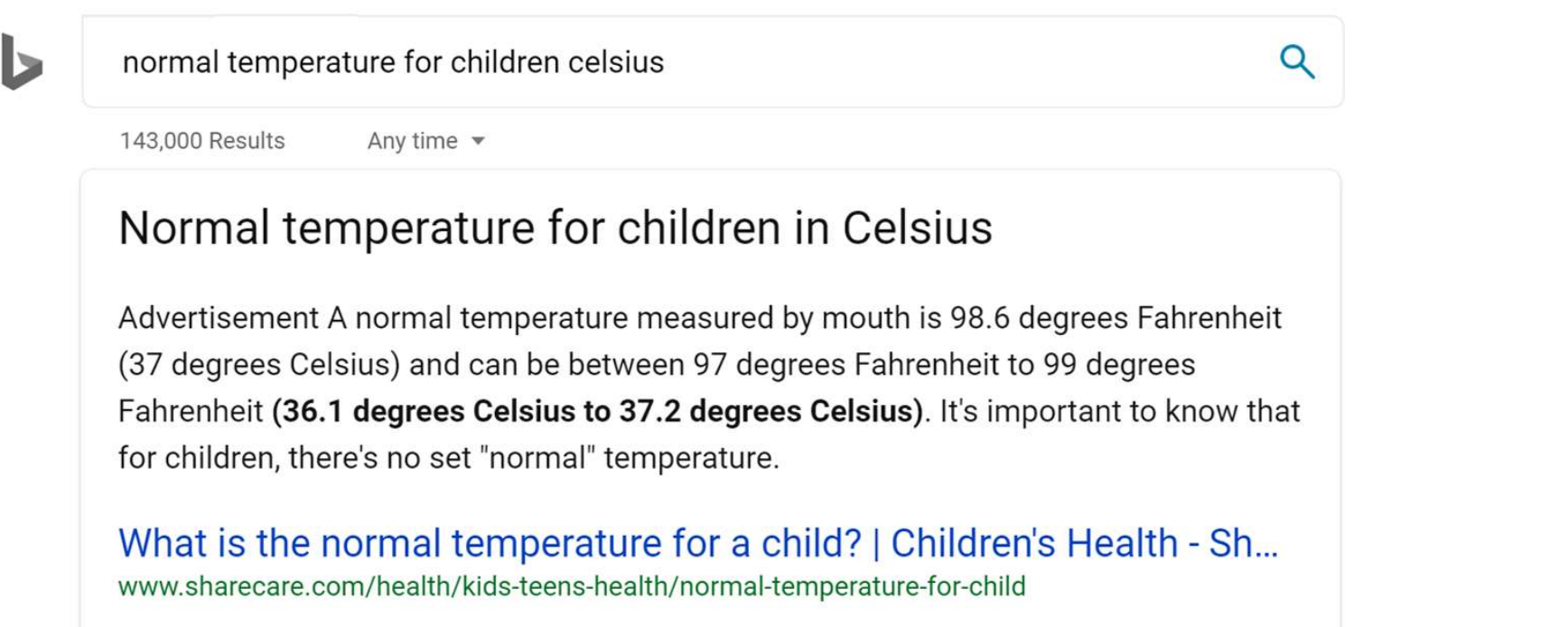}
\caption{Example QA features in web search engines.}
\label{Goole&Bing}
\end{figure}

Question answering (QA) has become a de facto feature in \emph{search result pages} (\emph{SERP} for short) in most commercial search engines. For a query bearing some question intent, such as a noun phrase like symptoms of coronavirus, a search engine can extract the most relevant passage from web documents and put it in an individual block at the top of a SERP.  Figure~\ref{Goole&Bing} shows the screenshot of the QA feature of a commercial search engine, where the query is ``normal temperature for children in Celsius''.

Typically, a {\em QA block} is composed of a question, the passage to answer the question, the URL of the source web document from which the passage is extracted, and the links to collect user feedback (e.g., ``Was the QA block helpful?''). Clearly, a well-designed QA block can deliver informative answers to search engine users in an intuitive and straightforward manner, save user time, and improve user experience. QA blocks have become even more popular on mobile devices as voice search is adopted by more and more users. 

No doubt the magic behind a QA block is empowered by various machine learning algorithms, including the latest deep neural networks~\cite{radev2002probabilistic,echihabi2008select,kaisser2004question,DBLP:journals/corr/abs-1908-06780,Yang2020ModelCW,Yuan2020EnhancingAB,Huang2019UnicoderAU}.  While machine learning algorithms have attracted extensive attention, people often overlook a critical challenge in making QA blocks industry scale commercial products -- the need of huge amounts of training data. In practice, a commercial search engine receives extremely diverse questions in open domain at web scale. To handle such a complex and huge question space, the QA models for search engines often have to involve tens of millions of parameters, which cause the models easily overfitting to small training data. Consequently, we usually have to use millions of training examples to train a model in order to overcome overfitting and biases.  

It is well recognized that obtaining large amounts of high quality training data is a bottleneck for commercial search engines. Using human judges to label training data is very expensive in financial cost and time.  To make the challenge even tougher, a commercial search engine often provides services in global markets with various languages. It is unrealistic to manually label millions of training samples for each language.  

A practical approach for search engines to collect massive training data for search tasks is to exploit implicit relevance feedback from user behavior mined from search logs.  There exists a rich body of literature~\cite{DBLP:journals/sigir/JoachimsGPHG17,Gao2011ClickthroughbasedLS,DBLP:conf/cikm/HuangHGDAH13,DBLP:journals/sigir/AgichteinBD18,Bilenko:2008:MST:1367497.1367505, White:2007:SUP:1277741.1277771}. Can we simply extend the existing best practice to collect user implicit relevance feedback data to train QA models?  Unfortunately, all existing works target at the relevance of web {\em documents}, rather than {\em passages}. Collecting and understanding implicit relevance feedback for QA blocks is much more sophisticated and demands dramatic innovation beyond any existing approaches.

While we will develop our novel approach and present our best engineering practice later in the paper, let us illustrate some challenges in collecting and understanding implicit relevance feedback for QA block using a real example.

Figure~\ref{t:passage_behavior} shows two QA examples. In the first case, ``{\sl What's the normal body temperature for \textbf{child}?}'', there is no click in the QA block. In the second case, ``{\sl What's the normal body temperature for \textbf{adult}?}'', user clicks on the URL in the QA block are observed.  We further examine these two cases, and find that the passage in the first case perfectly answers the question. Therefore, a user can obtain satisfactory information by simply going through the content of the passage. No follow-up action is needed. 

For the second case, the information in the passage (about child body temperature) does not accurately match the user intent (for adult body temperature). A user may have to explore more information in the source page from which the passage is extracted. In this case, the title ``human body temperature'' of the source page may trigger a user's interest to click on the URL and read more in that page. This example illustrates a unique characteristic of QA. As the content of the passage is already presented to users in a QA block, users may not need to click on the URL to get the answer. Consequently, the correlation between user clicks and passage relevance may be much weaker than the correlation between user clicks and page relevance in web search results. We will provide more insights in Section~\ref{sec:click_baseline}.

\begin{figure}[t]
    \small
    \centering
    \begin{tabular}{lp{5cm}p{7cm}}
    \hline
    \textbf{Question (a)}: &\emph{What's the normal body temperature for \textbf{child}?}\\
    \hline
    \textbf{Passage (a)}: &\emph{The average normal body temperature for children is about \textbf{37 degree}. A child's temperature usually averages from around \textbf{36.3 degree} in the morning to \textbf{37.6 degree} in the afternoon.} \\ \hline
    \textbf{URL (a)}: &\emph{Human Body Temperature: Fever - Normal - Low \url{https://www.disabled-world.com/calculators-charts/degrees.php}} \\
    \hline
    \textbf{Label}: &\emph{Relevant} \\
    \hline
    \textbf{User behavior}: &\emph{\textbf{No Click}} \\
    \hline
    \\
    \hline
    \textbf{Question (b)}: &\emph{What's the normal body temperature for \textbf{adult}?}\\
    \hline
    \textbf{Passage (b)}: &\emph{The average normal body temperature for children is about \textbf{37 degree}. A child's temperature usually averages from around \textbf{36.3 degree} in the morning to \textbf{37.6 degree} in the afternoon.} \\ \hline
    \textbf{URL (b)}: &\emph{Human Body Temperature: Fever - Normal - Low \url{https://www.disabled-world.com/calculators-charts/degrees.php}} \\
    \hline
    \textbf{Label}: &\emph{Irrelevant} \\
    \hline
    \textbf{User behavior}: &\emph{\textbf{Click}} \\
    \hline
    \end{tabular}
    \label{table:example}
    \caption{\label{t:passage_behavior}Examples of user behavior for web QA.}
    \vspace{-8pt}
\end{figure}


Another major difference between QA block and web document results is the number of results presented in SERP. Given a user question, a search engine usually returns a list of web documents, but only a single QA block. Most previous click models leverage the relative rank order of documents to gain more reliable implicit feedback. However, this idea cannot be used to QA blocks, since SERP contains only one QA block per question. 



In this paper, we investigate user behavior in QA blocks and propose a novel approach to mine implicit relevance feedback from noisy user behavior data for passage relevance. To the best of our knowledge, this is the first systematic study submitted for publication to address the data collection challenges for QA blocks. We make the following contributions.

First, we capture three types of user behavior when users interact with QA blocks, namely \emph{click behavior}, \emph{re-query behavior}, and \emph{browsing behavior}. By analyzing the aggregated sequences of user actions in the context of complete search sessions, we obtain interesting insights about the correlation between user behavior and passage relevance. 

Second, we examine several possible methods that automatically extract user feedback signals from user behavior data. With a small amount of human labeled data as ground truth, we reveal strong correlation between extracted feedback signals and passage relevance, and further assess the feasibility of learning implicit feedback with reasonable accuracy. 

Third, we incorporate implicit feedback mined from user behavior data into a weakly-supervised approach for QA model training, and carry out extensive experiments on several QA datasets in English. The experimental results clearly show our approach greatly improves the QA performance on all datasets, especially under the low-resource conditions.

Last, we deploy our approach in a commercial search engine in two non-English markets. We find users speaking different languages uniformly follow similar behavior patterns when they interact with QA blocks. Consequently, the implicit relevance feedback model trained in en-US (english) market can be successfully transferred to foreign markets without any tuning. In de-DE (German) and fr-FR (French) markets, our approach significantly improves the QA service by around 3.0\% in the AUC metric. Moreover, this approach can automatically refresh the QA model by continuously collecting relevance feedback from users, which further saves the labeling cost. We expect our approach can save millions of dollars of labeling cost when scaling out to more markets. 

The rest of the paper is organized as follows. We first review the related work in Section~\ref{sec:related-work}. We then present our approach in Section~\ref{sec:method}. We report the extensive experimental results in Section~\ref{sec:experiment}, and conclude the paper in Section~\ref{sec:conclusion}.

\section{Related Work}
\label{sec:related-work}

Our study is mainly related to the previous work on QA and learning from user feedback.  We provide a brief review on those topics here.

\subsection{Question Answering (QA)}

The purpose of web QA is to offer users an efficient information access mechanism by directly presenting an answer passage to the web search engine users~\cite{chen2017reading,ahn2004using,buscaldi2006mining}. 
%
%
There are various methods for web QA in literature. For example, Moldovan~\textit{et al.}~\cite{DBLP:conf/acl/MoldovanHPMGGR00} proposed a window-based word scoring technique to rank potential answer pieces for web QA. Cui~\textit{et al.}~\cite{DBLP:conf/sigir/CuiSLKC05} learned transformations of dependency path from questions to answers to improve passage ranking. Yao~\textit{et al.}~\cite{DBLP:conf/naacl/YaoDCC13} tried to fulfill the matching using minimal edit sequences between dependency parse trees. AskMSR~\cite{DBLP:conf/emnlp/BrillDB02}, a search-engine based QA system, used Bayesian Neural Network relying on data redundancy to find short answers.


In recent years, deep neural networks have achieved excellent performance in QA~\cite{chen2017reading,DBLP:conf/aaai/WangYGWKZCTZJ18}. Convolutional Neural Networks (CNN) and Recurrent Neural Networks (RNN), such as the Long Short-Term Memory (LSTM) and Gated Recurrent Unit (GRU), were applied to learn the representations of questions and answers~\cite{DBLP:journals/corr/TanXZ15, DBLP:conf/acl/TanSXZ16}. Attention mechanisms were employed to model the interaction between questions and answers~\cite{DBLP:conf/cikm/YangAGC16,DBLP:conf/ijcai/WangHF17}, which led to better performance than simply modeling query and answer separately. Most recently, deep pre-trained models, such as BERT~\cite{DBLP:journals/corr/abs-1908-08167}, XLNet~\cite{DBLP:journals/corr/abs-1906-08237}, have become the new state-of-the-art approaches of QA models. 

To tackle web-scale open-domain question answering, the statistic machine learning models require large amounts of training data. In this paper, we do not target at developing another QA model. Instead, we aim to find a model-agnostic approach for training data collection.

\subsection{Learning from User Feedback}
User feedback has been intensively explored in web page ranking to improve search quality~\cite{DBLP:conf/kdd/Joachims02, DBLP:conf/sigir/JoachimsGPHG05}. There are two types of user feedback. \emph{Explicit (or shallow) feedback} is that a user takes an extra effort to proactively express her satisfaction with the search results, e.g., through a simple up-voting or down-voting button. \emph{Implicit feedback} is the inference of user satisfaction from a user's search and/or browse sessions, without extra efforts from the users.

Rocchio~\cite{rocchio1971relevance} is a pioneer to leverage relevance feedback for information retrieval by explicitly gathering feedback through a button for up-voting or down-voting.  Another means of collecting explicit feedback was through side-by-side comparisons~\cite{ali2006relationship, thomas2006evaluation}. In practice, the chances of receiving explicit feedback from users are very low, since explicit feedback disturbs users in their normal interaction with search engines.

Compared with explicit feedback, implicit feedback can be collected at much lower cost and in much larger quantity, without putting any burden on users of search systems~\cite{DBLP:journals/sigir/JoachimsGPHG17}. Various features have been extracted from user behavior data, such as click-through information, average dwell time, and number of page visits in post-search browsing sessions~\cite{DBLP:journals/sigir/AgichteinBD18,Bilenko:2008:MST:1367497.1367505}.  For example, Joachims~\textit{et al.}~\cite{DBLP:journals/sigir/JoachimsGPHG17} derived relative preferences from click-through information. Agichtein~\textit{et al.}~\cite{DBLP:journals/sigir/AgichteinBD18} explored page clicks and page visits as features to improve ordering of top results in web search. Gao~\textit{et al.}~\cite{DBLP:conf/sigir/GaoTY11} and Huang~\textit{et al.}~\cite{DBLP:conf/cikm/HuangHGDAH13} used click-through data for deep semantic model training to learn semantic matching between queries and documents in page ranking. A major challenge in exploiting implicit feedback is that it is inherently noisy or even biased~\cite{DBLP:conf/sigir/JoachimsGPHG05}. To address the challenge, various methods have been proposed. For example, Craswell~\textit{et al.}~\cite{DBLP:conf/wsdm/CraswellZTR08} proposed four simple hypotheses about how position bias may arise. Dupret and Piwowarski~\cite{DBLP:conf/sigir/DupretP08} proposed a set of assumptions on user browsing behavior in order to estimate the probability a document was to be viewed. Chapelle and Zhang~\cite{DBLP:conf/www/ChapelleZ09} proposed a dynamic Bayesian Network model to indicate whether a user was satisfied with a clicked document and then left the page.

Although user feedback for web page ranking has been well studied, there is little work on user feedback for web QA. The closest work to our study is by Kratzwald and Feuerriegel~\cite{kratzwald2019learning}, who designed feedback buttons to explicitly ask users to assess the overall quality of the QA result. Different from them~\cite{kratzwald2019learning}, our work mainly focuses on the mining of implicit relevance feedback for web QA. To the best of our knowledge, it is the first study in this frontier.

\nop{
\begin{figure*}[ht]
\setlength{\belowcaptionskip}{-0.2cm}
\centering
\includegraphics[scale=0.68, viewport=20 180 700 470, clip=true]{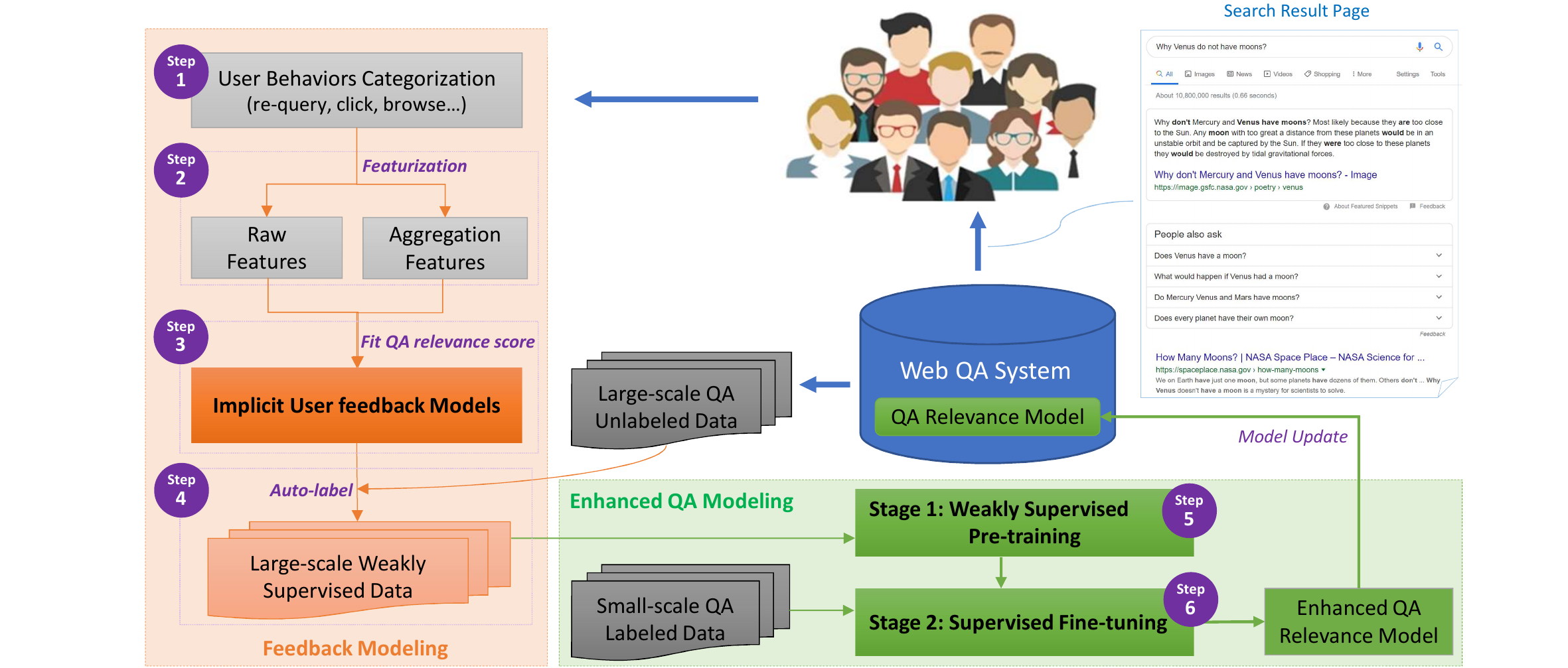}
\caption{\label{figure:3_framework} The overall framework of our FeedbackQA approach.}
\label{fig:framkework}
\end{figure*}
}

\section{Our Approach}
\label{sec:method}


Our goal is to derive training data from online user behavior to train QA models. To achieve this goal, our basic idea is to learn an implicit relevance feedback model. Given a question $q$ and the passage $p$ served by the search engine, the feedback model extracts features from user behavior and predicts the relevance $r$ between $q$ and $p$. The predicted results $(q, p, r)$ are then used as training data to train QA models. 

Based on this idea, we first conduct a comprehensive analysis on user behavior in QA sessions in Section~\ref{sec:user_feedback} and propose a systematic categorization to cover all types of user behavior. We then design a rich set of user behavior features in Section~\ref{sec:feature_extraction} to make sure we do not miss any useful implicit feedback signals. In Section~\ref{sec:imp_feedback_modeling}, we carefully compare various algorithms that learn implicit feedback models, and apply the best model to a huge volume of user behavior data to derive a large amount of training data. Section~\ref{sec:two_stage} elaborates how we leverage the derived training data in a weakly-supervised approach for the training of QA model.

\subsection{Taxonomy of User Behavior}
\label{sec:user_feedback}

\begin{table}[t]
\caption{\label{table:behavior} Taxonomy of user behavior in web QA system}
\begin{center}
\begin{tabular}{ ccc} 
 \hline
 \textbf{Type} & \textbf{Type} & \textbf{behavior} \\
 \hline
  \textbf{Explicit}  & \emph{Click} & \emph{Up-vote/Down-vote} \\
 \hline
     \multirow{8}{*}{\textbf{Implicit}} & \emph{Re-query} & \emph{Reformulation} \\
 \cdashline{2-3}
    & \multirow{4}{*}{\emph{Click}} & \emph{Answer Click} \\ 
    &   & \emph{Answer Expansion Click} \\ 
    &   & \emph{Outside Answer Click} \\ 
    &   & \emph{Related Click} \\ 
 \cdashline{2-3}
    & \multirow{1}{*}{\emph{Browsing}} & \emph{Browse} \\
 \hline
\end{tabular}
\label{table:user-feedback}
\end{center}
\vspace{-10pt}
\end{table}

We propose a taxonomy of user behavior summarized in Table~\ref{table:behavior}.  At the higher level, we distinguish two types of user behavior, which correspond to explicit and implicit feedback to web QA systems. In the following, we first show an empirical study of explicit feedback in a commercial search engine, and explain why it is not efficient or effective to collect training data through explicit feedback. We then describe user implicit feedback in detail.

To collect explicit feedback, commercial search engines, such as Google and Bing, provide links at the bottom of a QA block, as illustrated in Figure~\ref{Goole&Bing}. However, only a very small fraction of users send their explicit feedback. In a real commercial web QA system, the coverage of explicit feedback, i.e., clicking on the feedback links, is only less than \emph{0.001\%} of the total QA impressions. Moreover, we find that users strongly tend to send negative feedback -- the positive over negative ratio is about 1:17. To form a balanced training dataset, we have to sample almost equal amounts of positive and negative examples from the skewed label distribution. This further reduces the size of valid training data that can be derived from explicit feedback. Consequently, the explicit feedback may not be a good source to collect training data for web QA model.

We then cast our attention to implicit feedback. Basically, all actions related to QA blocks recorded in search logs are either queries and clicks. Therefore, the first two categories in our taxonomy correspond to these two types of actions. We further refine the categories of actions related to QA blocks into sub-groups. For example, we distinguish the types of clicks based on the components that are clicked on. Finally, we also model the information about general SERP browsing that may be useful to QA blocks, which is the last category in our taxonomy. The details of the taxonomy are introduced in the following.

\noindent 
\textbf{\emph{Re-query Behavior}}: we consider the sequence of user queries in a session and particularly note whether a user issues a new query by modifying the previous one in the session. Interchangeably we also call this behavior \emph{reformulation}.

\begin{figure}[t]
\setlength{\belowcaptionskip}{-0.2cm}
\centering
\includegraphics[scale=0.78, viewport=250 160 480 462, clip=true]
{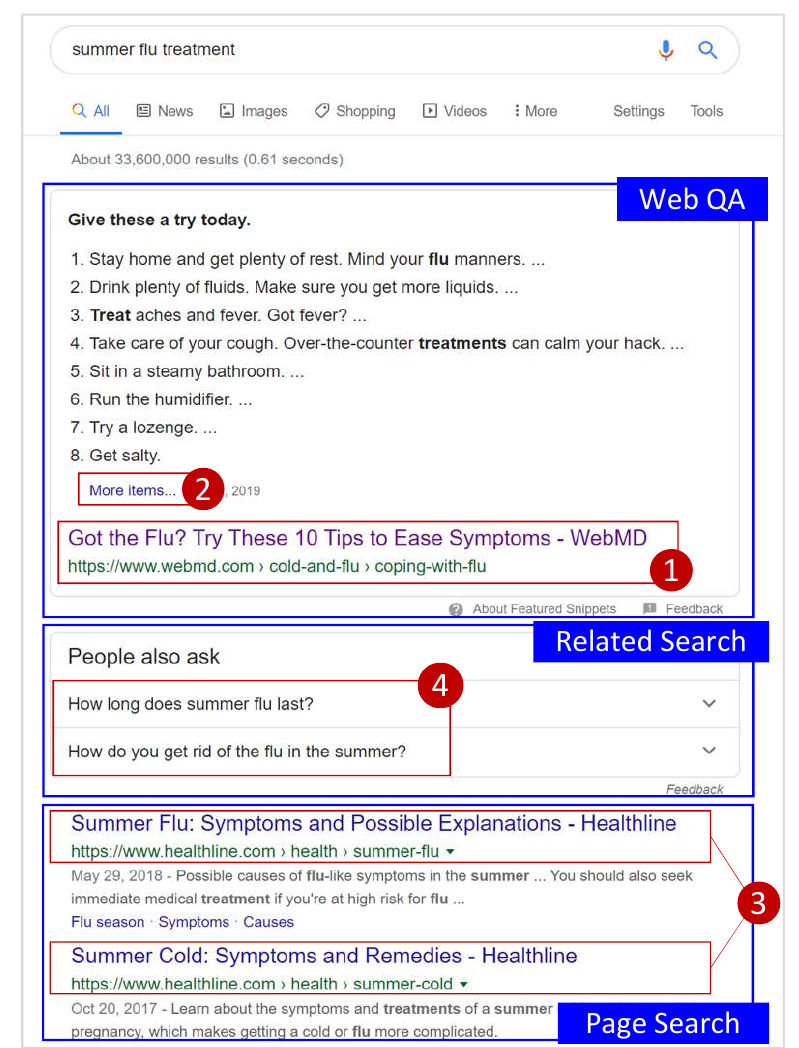}
\caption{\label{figure:user_behavior} Illustration of user click behavior, including \emph{Answer Click}, \emph{Answer Expansion Click}, \emph{Outside Answer Click}, and \emph{Related Click}.}
\label{fig:impact_size}
\end{figure}

\noindent
\textbf{\emph{Click Behavior}}:  we distinguish four types of clicks, depending on the components being clicked on (see Figure~\ref{figure:user_behavior} for illustration). 
\begin{itemize}
    \item \emph{Answer Click}: a user clicks on the source page URL of the answer passage (indicated by {\small \textcircled{1}} in Figure~\ref{figure:user_behavior}).
    \item \emph{Answer Expansion Click}: a user clicks on a special button (indicated by {\small \textcircled{2}} in Figure~\ref{figure:user_behavior}) to expand the folded QA answer due to the maximum length limit for display. 
    \item \emph{Outside Answer Click}: a user clicks on the links to the web documents in the SERP (indicated by {\small \textcircled{3}} in Figure~\ref{figure:user_behavior}) other than the source page URL for the web QA passage. 
    \item \emph{Related Click}: a user clicks on the related queries (indicated by {\small \textcircled{4}} in Figure~\ref{figure:user_behavior}) to explore more information.
\end{itemize}

\noindent
\textbf{\emph{Browsing Behavior}}: a user reads the content of the QA passage or other components in the SERP, and does not give any input to the search engine.

\subsection{Feature Extraction from User Behavior}\label{sec:feature_extraction}
\begin{table}[t]
\small
\caption{\label{table:feature_description} User behavior features. Here ``Answer'' means the source URL for the passage in web QA.}
\begin{tabular}{p{2.0cm}lp{4.3cm}}
\hline
\textbf{Name}     & \textbf{Type}     & \textbf{Description} \\
\hline 
\emph{RFRate}         &  \emph{Re-query}  &  rate of re-query     \\
\hdashline
\emph{AnswerCTR}         &  \emph{Click}     &  CTR of answer      \\
\emph{AnswerOnlyCTR}     &  \emph{Click}     &  CTR with only click on answer      \\
\emph{AnswerSatCTR}      &  \emph{Click}     &  satisfied CTR of answer\\
\emph{AnswerExpRate}      &  \emph{Click}     &  CTR of answer expansion                          \\
\emph{OTAnswerCTR}       &  \emph{Click}     &  CTR outside of answer                           \\
\emph{OTAnswerOnlyCTR}    &  \emph{Click}     &  CTR with only click outside of answer                           \\
\emph{OTAnswerSatCTR}    &  \emph{Click}     &  satisfied CTR outside of answer                           \\
\emph{BothClickCTR}    &  \emph{Click}     &  CTR of both click on/outside of answer                     \\
\emph{RelatedClickRate}  &  \emph{Click}     &  CTR of related queries                            \\

\hdashline
\emph{NoClickRate}       &  \emph{Browsing}     &  no click rate               \\
\emph{AbandonRate}       &  \emph{Browsing}  &  abandonment rate                        \\
\emph{AvgSourcePage- DwellTime}  &  \emph{Browsing}  & average source page dwell time                  \\
\emph{AvgSERPDwellTime}  &  \emph{Browsing}  & average SERP dwell time                  \\
\hline
\end{tabular}
\end{table}

To learn implicit feedback models, we need to design user behavior features, which should be sensitive and robust in capturing relevance feedback signals from users. To meet this goal, we follow two principles in our feature design. First, our feature set exhaustively covers all types of user behavior discussed in Section~\ref{sec:user_feedback}, and makes sure we do not miss any useful relevance signals. Second, we design aggregated features that summarize the behavior of multiple users in multiple sessions. Aggregated features can effectively reduce noise and biases in individual users and individual actions. The features are listed in Table~\ref{table:feature_description}. Most features are straightforward. Please refer to the ``Description'' column for the meaning of the features. We only select several features to explain as follows.
\begin{itemize}
    \item \emph{AvgSourcePageDwellTime}: the average time from the user clicking into the source page of the web QA answer to the the user leaving the source page. 
    
    \item \emph{AvgSERPDwellTime}: the average time from SERP loaded successfully to the completion of the search session.  
    
    \item \emph{AbandonRate}: the percentage of sessions with no click on the SERP. In those sessions, a user just browses SERP and leaves the search sessions. 
\end{itemize}

The click-through rate (\emph{CTR}) for a component, which can be either a QA passage, an answer expansion, a related search, or a web document outside of QA block, is defined as 
\begin{equation}
CTR =  \frac{N_{click}}{N_{impression}}
\end{equation}
where $N_{impression}$ denotes the total number of impressions of the component and $N_{click}$ the number of clicks on the component. 

A satisfied click (\emph{SatClick}) on a component is a click on the component followed by the corresponding Dwell Time greater than or equal to a pre-defined threshold. The satisfied click-through rate (\emph{SatCTR}) is then defined as 
\begin{equation}
 SatCTR =  \frac{N_{SatClick}}{N_{impression}}  
\end{equation}
 where $N_{SatClick}$ denotes the number of SatClicks on the component.

\subsection{Implicit Feedback Modeling}
\label{sec:imp_feedback_modeling}
In this section, we target at building an effective implicit feedback model to predict the relevance between a question and a passage based on the features designed in Section~\ref{sec:feature_extraction}. We first prepare a dataset with 18k QA pairs, where each QA pair is augmented with the set of user behavior features as well as a human judged label (Section~\ref{sec:feedback_dataset}). Intuitively, a single feature, such as AnswerCTR or AnswerSatCTR, may be too weak a signal to infer user satisfaction. We verify this assumption as a baseline in Section~\ref{sec:click_baseline}. We then consider various machine learning models, including Logistic Regression (LR)~\cite{menard2002applied}, Decision Tree (DT)~\cite{safavian1991survey}, Random Forest (RT)~\cite{liaw2002classification}, and Gradient Boost Decision Tree (GBDT)~\cite{ke2017lightgbm}, and conduct an empirical study to choose the best model (Section~\ref{sec:ml_models}). We also analyze the feature importance and derive rules from the learned models, which help us gain insights into users' decision process when they interact with a QA block.

\subsubsection{Dataset}
\label{sec:feedback_dataset}
We create a dataset which consists of 18k QA pairs, where each QA pair is augmented with the set of user behavior features as well as a human judged label. To be more specific, the dataset is a table, where each row is in the form of \emph{$\langle$Question, Passage, User behavior features, Label$\rangle$}. The QA pairs are sampled from a half-year log from January to June 2019) of one commercial search engine. We sample the QA pairs whose number of impressions in the log is around 50. This number is set due to two considerations. First, we want to aggregate multiple users' behavior to reduce the noise from individual users. Second, we want to avoid too popular queries, which tend to be short and too easy to answer. Each QA pair is sent to three crowd sourcing judges and the final label is derived based on a majority voting. This 18k dataset is further randomly split into 14k/2k/2k as training, dev and test sets, respectively. We plan to make this data set public if this paper is accepted.

\subsubsection{Baseline}\label{sec:click_baseline}
\begin{figure}[t]
\setlength{\belowcaptionskip}{-0.2cm}
\centering
\includegraphics[scale=0.32, viewport=10 10 700 350, clip=true]{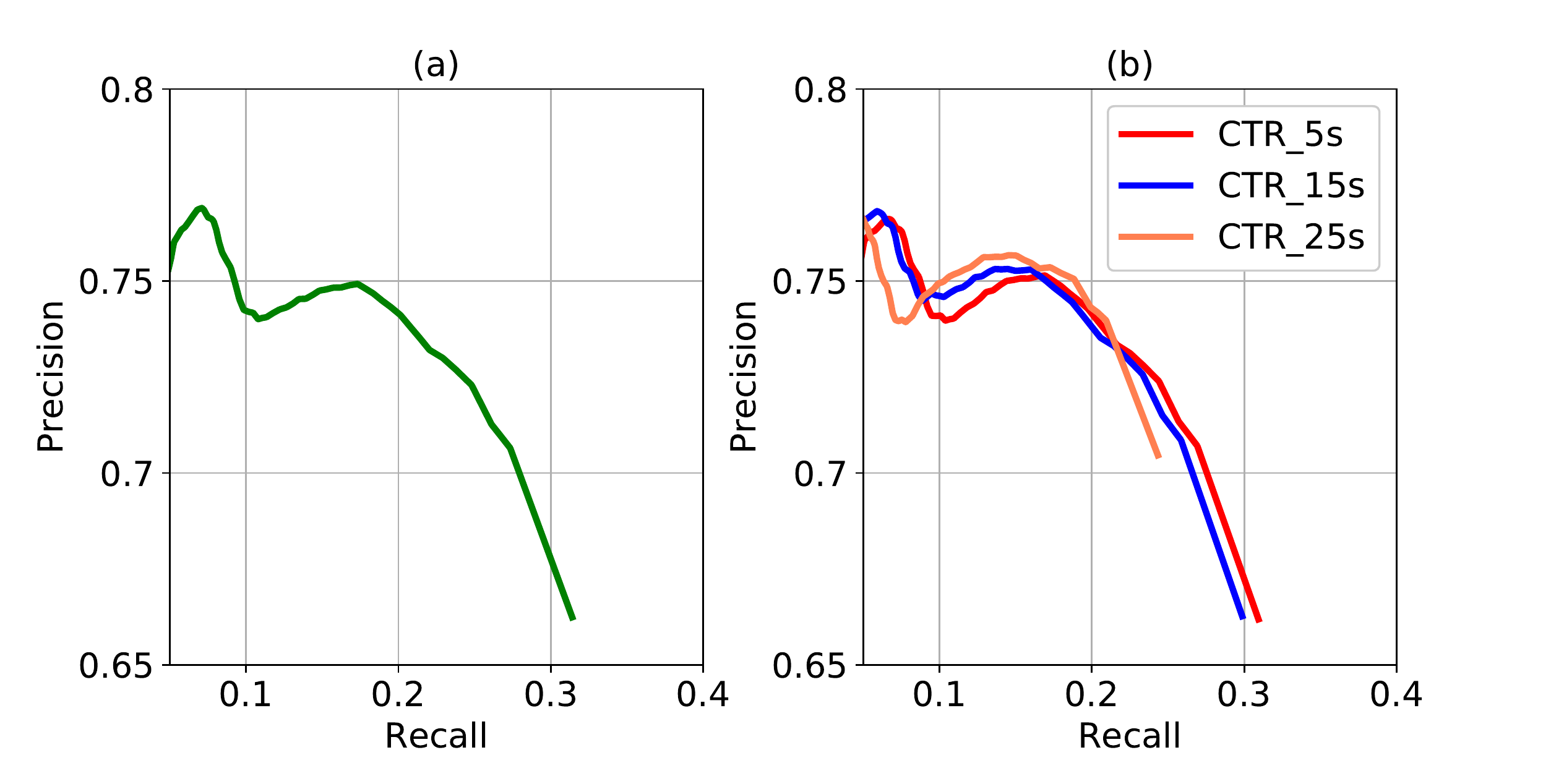}
\caption{\label{figure:pr}Precision-recall curves of \emph{AnswerCTR}(a) and \emph{AnswerSatCTR}(b).}
\label{PR_Curve}
\vspace{-5pt}
\end{figure}

Click-through rate (\emph{CTR}) and satisfied click-through rate (\emph{SatCTR}) have been widely adopted in existing works as indicators for the relevance of a web page with respect to a given user query. Analogously in our study of passage relevance, we start with users' clicks on the source URL of the answer passage. We first investigate the feature of $AnswerCTR$ in Table~\ref{table:feature_description} by plotting a precision-recall curve in Figure~\ref{figure:pr}(a).  
\begin{itemize}
\item For QA pairs whose $CTR$ > 0, the recall is less than 0.33. In other words, when the passage is relevant to the question, there are more than two thirds of cases where users do not make any single click on the source URL. Please note that the number of clicks is counted throughout all the impressions for that question-passage pair. This is a very different observation compared to the cases for page ranking. However, when we consider the feature of question answering, this result is not surprising since users may simply browse the content of the passage and get the information. No further click into the source URL is needed at all.
\item The highest precision is less than 0.77, across the full range of recall values. This suggests that clicking into the source URL does not necessarily suggest a relevant passage. We find in most clicked cases, the passage is partially relevant to the question. Therefore, users may want to click into the source URL to explore more information.
\end{itemize} 

We further investigate the correlation between \emph{SatCTR} and passage relevance. Similarly, we plot the precision-recall curves in Figure~\ref{figure:pr}(b). $CTR\_t$ means we consider the clicks followed by dwell time on the source page for longer than $t$ seconds as a satisfied click. We experiment with dwell time thresholds 5s, 15s, and 25s, and observe similar trend as in Figure~\ref{figure:pr}(a).

The experiments with \emph{CTR} and \emph{SatCTR} verify that the single feature of user clicks into the source URL is not a good indicator for passage relevance. Clicks do not necessarily indicate relevant passages, and vice versa. Thus, we have to consider more complex models to combine the sequences of user actions in search sessions.

\begin{table}
\begin{center}
\small
\caption{Comparison of feedback modeling methods. The best results are highlighted in bold, and the secondary best results are marked with underline.}
\begin{tabular}{llll}
\hline
\textbf{Method} & \textbf{AUC} & \textbf{ACC} & \textbf{F1} \\
\hline
\emph{AnswerCTR}                       & 58.28         & 53.87         & 18.11 \\
\emph{AnswerSatCTR\textsubscript{5s}}  & 57.73 (-0.55) & 53.50 (-0.37) & 16.52 (-1.59) \\
\emph{AnswerSatCTR\textsubscript{15s}} & 57.75 (-0.53) & 52.73 (-1.14) & 13.43 (-4.68)  \\
\emph{AnswerSatCTR\textsubscript{25s}} & 57.79 (-0.49) & 52.63 (-1.24) & 12.23 (-5.88) \\
\hline
\textbf{LR}          & \underline{71.41 (+13.13)} & 66.00 (+12.13) & {\bf 66.78 (+48.67)} \\
\textbf{DT}          & 61.75 (+3.37)  & 63.43 (+9.56)  & 60.92 (+42.81) \\
\textbf{RF}          & 71.14 (+12.86) & \underline{67.47 (+13.60)} & 65.66 (+47.55) \\
\textbf{GBDT}        & \textbf{\textbf{73.69 (+15.41)}} & \textbf{\textbf{68.00 (+14.13)}} & \underline{66.08 (+47.97)} \\
\hline 
\end{tabular}
\label{table:metrics for different modeling method}
\end{center}
\end{table}

\subsubsection{Machine Learning Models}\label{sec:ml_models}
We apply machine learning models to combine the various types of user behavior features. The training target is to fit the human judged labels. We evaluate the model performance by common binary classification metrics, including area under the curve (AUC), accuracy (ACC), and F1 score. 

\noindent 
\textbf{\emph{Machine Learning Models Considered}}. We apply various models and evaluate the results, including Logistic Regression (LR), Decision Tree (DT), Random Forest (RT), and Gradient Boost Decision Tree (GBDT). 

\noindent 
\textbf{\emph{Results}}. As summarized in Table~\ref{table:metrics for different modeling method}, the machine learning approach significantly outperforms baseline methods (i.e. \emph{AnswerCTR} and \emph{AnswerSatCTR}) on all metrics. In terms of AUC and ACC, the GBDT model achieves the best performance. In terms of F1 score, the performance of GBDT model (66.08) is very close to the best result (66.78). Overall, we consider GBDT as the best model.

\begin{figure}[t]
\setlength{\belowcaptionskip}{-0.2cm}
\centering
\includegraphics[scale=0.45, viewport=10 10 450 230, clip=true]{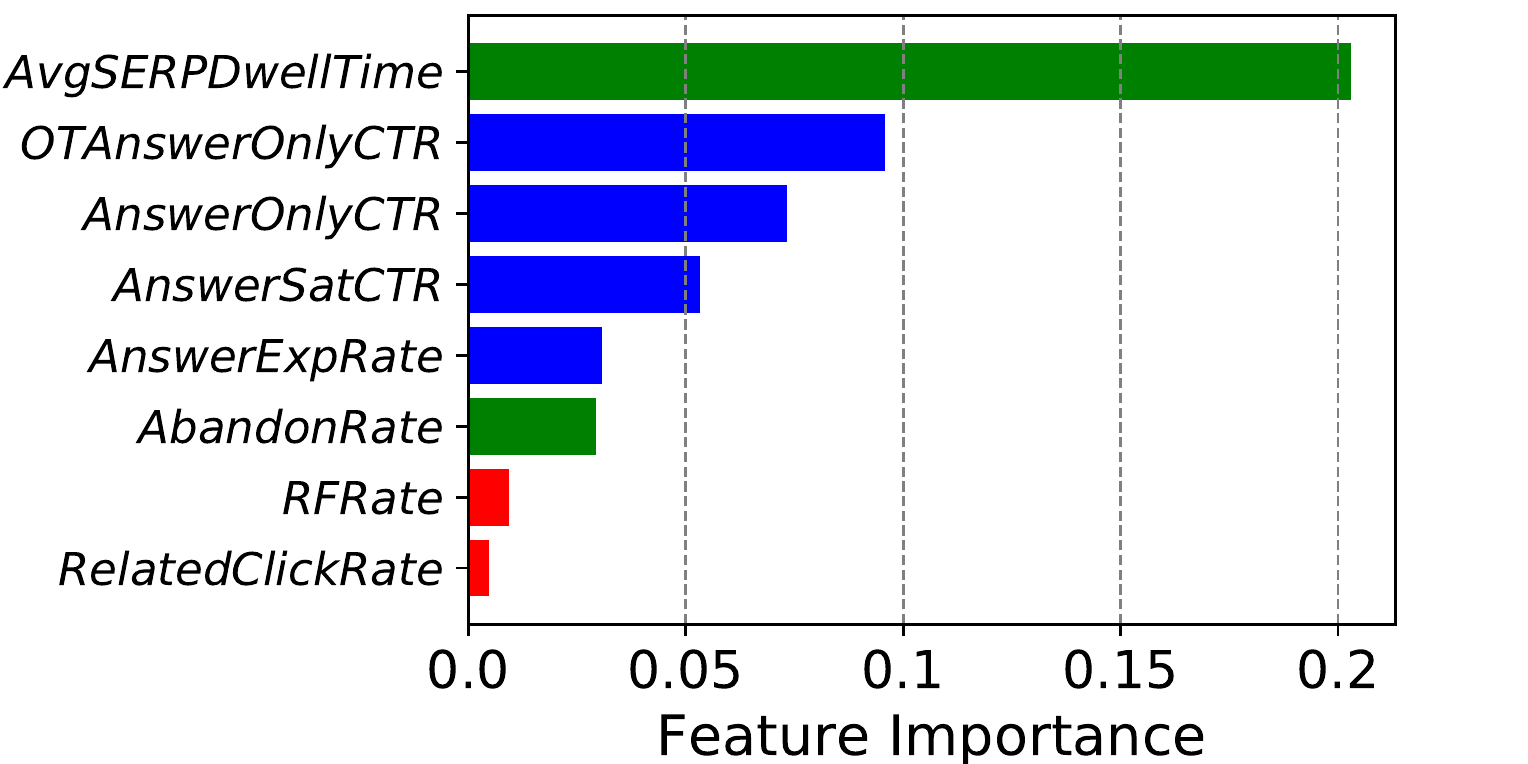}
\caption{\label{figure:Feature_importance} Relative weights of top 8 features in the GBDT model.}
\label{fig:impact_size}
\vspace{-5pt}
\end{figure}

\noindent 
\textbf{\emph{Model Interpretation}}. 
To get more insights about user behavior on QA, we first investigate the impact of individual feature based on the best model GBDT. The top 8 features of the model are shown in Figure~\ref{figure:Feature_importance}. 

The results indicate that \emph{AvgSERPDwellTime} and \emph{OTAnswerOnlyCTR} have the highest feature importance, followed by \emph{AnswerOnlyCTR}, \emph{AnswerSatCTR}, \emph{AnswerExpRate} and \emph{AbandonRate}. Reformulation related features like \emph{RFRate} as well as \emph{RelatedClickRate} have relatively low importance.  We can also obtain the following insights about user behavior on web QA.
\begin{itemize}
    \item Click features related to web QA answer itself, such as \emph{AnswerOnlyCTR} and \emph{AnswerSatCTR}, are not the most important features. This aligns well with our previous observations. 
    \item  SERP Dwell Time suggests the time period during which a user lingers on the search result page. Since the content of passage is presented in the QA block (in the SERP) as the answer to the user's question, the length of SERP Dwell time may be a good indicator of the relevance of the passage. 
\end{itemize}

To further reveal users' decision process in a search session, we examine the decision tree model DT in Table~\ref{table:metrics for different modeling method}. Some interesting insights gained from the paths on the tree are listed in the following.
\begin{itemize}
    \item If ``\emph{AvgSERPDwellTime} is long $\wedge$ \emph{NoClickRate} is large'', i.e., the SERP is abandoned, the passage usually has good relevance. Users may just browse the QA passage for a while and then get the information needed. 
    \item ``\emph{AnswerCTR} is small $\wedge$ \emph{OTAnswerCTR} is large'' is often a strong signal that the passage has poor relevance. In such cases, users may not be satisfied with the passage answer, and then click more on other documents.
    \item ``\emph{AnswerOnlyCTR} is large $\wedge$ \emph{AvgSERPDwellTime} is long'' is also a positive signal for passage relevance. The passage is relevant to the question, but due to the space limit of the QA block, the displayed content cannot fully answer user's question. Therefore, the user clicks on the source URL.
    \item ``\emph{NoClickRate} is large $\wedge$ \emph{RelatedClickRate} is large'' suggests that the passage is not relevant to the user question. The user revises the query to express her information need.
\end{itemize}

\noindent 
\textbf{\emph{Summary}}: Unlike the cases of page relevance, user clicks (including satisfied clicks) are not a good indicator for passage relevance in web QA. However, using machine learning models to combine various user behavior, it is still feasible to extract relevance feedback from the search sessions.

\subsection{Pre-training QA models} \label{sec:two_stage}
Through an implicit feedback model, we can derive a large amount of training data. However, user behavior data may be very noisy. Although we develop aggregated features to reduce biases in individual sessions, the prediction accuracy of the best model is only 68\% (see Table~\ref{table:metrics for different modeling method}).  How to leverage such noisy training data becomes an interesting problem. 

Inspired by the pre-training idea in learning deep neural networks, we apply large-scale automatically derived implicit feedback data to pre-train QA models, such as BiLSTM and BERT, as weak supervision. Intuitively, although the derived training data is noisy, it still contains valuable relevance signals and can roughly guide the parameters of QA models to a region close to the optimal solutions. In the second stage, we apply the human labeled data to fine-tune the parameters to get the final model. As verified in our experiments (Section~\ref{sec:experiment}), the strategy of pre-training plus fine-tuning is remarkably better than training models with only human labeled data without pre-training.

\nop{
As shown in Figure~\ref{fig:framkework}, a two-stage approach is proposed to integrate the user implicit feedback signals to the QA relevance model training process. At the first stage, a huge number of tuples \emph{<Query, Passage, User feedback>} can be mined from the search logs of a commercial search engine. By applying the feedback models developed in Table~\ref{table:metrics for different modeling method}, each \emph{<Query, Passage>} pair ($\left \langle Q, P \right \rangle$) is auto-labeled with a predicted relevance score. 
\begin{align}
&score_{<Q,P>}=F_{FeedbackModel}(x_{1}, ..., x_{m}) \\
&label_{<Q,P>}=\left\{\begin{matrix}
1 & score_{<Q,P>} \geq \tau_1,\\ 
0 & score_{<Q,P>} \leq \tau_2
\end{matrix}\label{eq:label}\right.
\end{align}
where $F_{FeedbackModel}(\cdot)$ denote the Feedback models, $x_{i}$ represents the feature described in the above section, $m$ represents the number of features, $\tau_1$ and $\tau_2$ are two thresholds.
}

Technically, let $Q={\{w_1, w_2, w_3, ..., w_{|Q|}\}}$ be a question with $|Q|$ words (or word pieces), $P={\{w_1, w_2, w_3, ..., w_{|P|}\}}$ be a passage with $|P|$ words (or word pieces). We use cross-entropy (CE) as our loss function for both pre-training and fine-tuning, defined by
\begin{align}
&y=F_{QAModel}(\left \langle Q, P \right \rangle) \\
&L_{CE}=-\frac{1}{n}\sum_{i}^{n}[\hat{y}_{i}\log(y_{i})+ (1-\hat{y}_{i})\log(1-y_{i})] \label{eq:cross_entropy}
\end{align}
where $F_{QAModel}(\cdot)$ are the QA models, which will be described in Section 4.2,  ${y}_{i}$ represents our QA relevance model output, $\hat{y}_{i}$ represents the true label, and $n$ is the number of training samples.

\section{Experiments}
\label{sec:experiment}

In this section, we report extensive experiments to verify our proposed approach using real data in a commercial search engine. 

\subsection{Datasets and Metrics}\label{sec:exp_data}
We conduct experiments on several datasets as follows, with their statistics shown in Table~\ref{t:statistic}. More detailed description of the following datasets as well as one extra dataset is presented in Appendix~\ref{sec:append_data}. AUC and ACC are used as the evaluation metrics for the QA relevance models.

\noindent \textbf{FeedbackQA\textsubscript{log}}: An English QA dataset collected from the latest half year's web QA system log in a commercial search engine. Each item of the log consists of a tuple $\langle$query, passage, user behavior$\rangle$. 
    
\noindent \textbf{FeedbackQA\textsubscript{\{ctr, gbdt\}}}: For each QA pair in FeedbackQA\textsubscript{log}, the feedback models in Table~\ref{table:metrics for different modeling method} are employed to predict a relevance label. The subscript indicates the model employed.

\noindent \textbf{DeepQA}: An English QA dataset where each case consists of three parts, i.e. question, passage, and a binary label (i.e. 0 or 1) by crowd sourcing human judges. The queries and passages are sampled from a commercial search engine. 

    
    
\noindent \textbf{MS Marco}: An open source QA dataset~\cite{DBLP:conf/nips/NguyenRSGTMD16}, which contains questions generated from real anonymized Bing user queries. To evaluate the effectiveness of our approach under low-resource setting (i.e., only a small amount of labeled data are available), the dataset is sub-sampled to form a positive/negative balanced set with 10k/1k/1k as training, dev and testing sets, respectively.

    
\noindent \textbf{WikiPassageQA}: A Wikipedia based open source QA dataset~\cite{DBLP:conf/sigir/CohenYC18}, targeting non-factoid passage retrieval tasks. To evaluate our approach in low resource setting, the dataset is sub-sampled to form a positive/negative balanced set with 10k/1k/1k as training, dev and testing sets, respectively.
    
\noindent \textbf{FrenchGermanQA}: This dataset is collected in a similar process as DeepQA. The main difference is that this dataset targets at French and German languages.

\begin{table}
\begin{center}
\small
\caption{\label{t:statistic} Statistics of experiment datasets.}
\begin{tabular}{lcccc}
\hline
\textbf{Dataset}          & \textbf{Train} & \textbf{Dev} & \textbf{Test} & \textbf{Labels}\\ 
\hline
\textbf{FeedbackQA\textsubscript{log}}               & 22M   & -    & -    & -             \\
\textbf{FeedbackQA\textsubscript{\{ctr, gbdt\}}} & 4M    & 10k  & 10k   & 50\%+/50\%-   \\
\hdashline
\textbf{DeepQA}               & 30k   & 2k   & 2k    & 57.6\%+/42.4\%-          \\
\textbf{MS Marco} &    10k         &      1k      &     1k  & 50\%+/50\%-                    \\ 
\textbf{WikiPassageQA} &    10k         &      1k      &     1k  & 50\%+/50\%- \\
\textbf{FrenchGermanQA} &    50k         &      2k      &     2k  & 50\%+/50\%- \\
\hline
\end{tabular}
\label{dataset}
\end{center}
\vspace{-8pt}
\end{table}

\begin{table*}[t!]
\small
\caption{\label{table:result}Performance comparison between our methods and baselines on the DeepQA dataset. All ACC and AUC metrics in the table are in percentage , where the sign \% are omitted.}\label{t:main}

\begin{tabular}{cccllll}
\hline
\multirow{2}{*}{\textbf{Model}}                    & \multirow{2}{*}{\textbf{Method}}                 & \multirow{2}{*}{ \textbf{\begin{tabular}[c]{@{}c@{}}Pre-training\\ Data Size\end{tabular}}} & \multicolumn{4}{c}{\textbf{Performance on Different Fine-tuning Data Size (AUC/ACC)}}                                                                  \\
\textbf{}                         & \textbf{}                        & \textbf{}       & \textbf{5k} & \textbf{10k} & \textbf{20k} & \textbf{30k}               \\ \hline

\multirow{7}{*}{\textbf{BiLSTM}}  & \multirow{1}{*}{\textbf{Original}} & \textbf{-}   & 60.45/58.21                        & 61.30/59.92                         & 61.55/61.99     &62.40/61.74 \\  \cdashline{3-7} 
                                  
                                  & \multirow{3}{*}{\textbf{FBQA\textsubscript{ctr}}} 
                                                                     & \textbf{0.5m}  & 59.90/57.60 (-0.55/-0.61)           & 61.25/58.25 (-0.05/-1.67)            & 61.40/60.50 (-0.15/-1.49)    
                                  & 60.65/59.29 (-1.75/-2.45) \\
                                  &                                  & \textbf{1.0m}    & 60.25/58.45 (-0.20/-0.24)           & 61.35/58.12 (+0.05/-1.80)            & 62.65/57.43 (+1.10/-4.56)       
                                  & 61.35/60.69 (-1.05/-1.05) \\
                                  &                                  & \textbf{4.0m}    & 60.50/56.99 (+0.05/-1.22)           & 59.75/58.39 (-1.55/-1.53)            & 60.90/59.15 (-0.65/-2.84)       
                                  & 62.25/61.73 (-0.15/-0.01) \\ \cdashline{3-7}
                                  
                                  & \multirow{3}{*}{\textbf{FBQA\textsubscript{FA}}} 
                                                                     & \textbf{0.5m}  & 61.95/59.66 (+1.50/+1.45)           & 62.50/60.96 (+1.20/+1.04)            & 62.85/62.74 (+1.30/+0.75)        
                                  & 64.23/62.50 (+1.83/+0.76) \\
                                  &                                  & \textbf{1.0m}    & 62.80/60.44 (+2.35/+2.23)           & 63.20/61.20 (+1.90/+1.28)            & 63.45/63.00 (+1.90/+1.01)       
                                  & 65.57/63.05 (+3.17/+1.31) \\
                                  &                                  & \textbf{4.0m}    & \textbf{64.13/62.15 (+3.68/+3.94)}           & \textbf{65.45/63.33 (+4.15/+3.41)}            & \textbf{65.46/64.17 (+3.91/+2.18)}       
                                  & \textbf{67.35/64.35 (+4.95/+2.61)} \\
                                  
\hline 

\multirow{7}{*}{\textbf{BERT}}    & \multirow{1}{*}{\textbf{Original}} & \textbf{-}   & 69.31/64.86                   &  71.81/67.76                  & 72.47/67.07  & 75.28/68.26 \\ \cdashline{3-7}

                                  & \multirow{3}{*}{\textbf{FBQA\textsubscript{ctr}}} 
                                                                     & \textbf{0.5m}  & 67.35/62.76 (-1.96/-2.10)      & 72.96/66.66 (+1.15/-1.10)      & 75.11/68.26 (+2.64/+1.19)   
                                  & 77.76/71.07 (+2.48/+2.81) \\
                                  &                                  & \textbf{1.0m}    & 72.33/67.06 (+3.02/+2.20)      & 73.76/67.36 (+1.95/-0.40)      & 76.16/69.16 (+3.69/+2.09)       
                                  & 77.42/68.26 (+2.14/+0.00) \\
                                  &                                  & \textbf{4.0m}    & 72.19/65.66 (+2.88/+2.90)      & 73.92/67.96 (+2.11/+0.20)      & 76.81/67.96 (+4.34/+0.89)      
                                  & 77.94/69.36 (+2.66/+1.10) \\ \cdashline{3-7}
                                  
                                  & \multirow{3}{*}{\textbf{FBQA\textsubscript{FA}}} 
                                                                     & \textbf{0.5m}  & 72.26/65.27 (+2.95/+0.41)      & 76.03/68.87 (+4.22/+1.11)      & 77.79/69.47 (+5.32/+2.40)   
                                  & 77.92/69.47 (+2.34/+1.21) \\
                                  
                                  &                                  & \textbf{1.0m}    & 73.53/66.37 (+4.22/+1.51)      & 76.29/68.97 (+4.48/+1.15)      & 78.63/68.77 (+6.16/+1.70)       
                                  & 79.82/70.17 (+4.54/+1.91) \\
                                  
                                  &                                  & \textbf{4.0m}    & \textbf{76.53/68.57 (+7.22/+3.71)}  & \textbf{78.17/68.57 (+6.36/+0.81)}   & \textbf{79.79/71.17 (+7.32/+4.10)}       
                                  & \textbf{81.03/71.57 (+5.78/+3.31)} \\

              \hline
\end{tabular}
\end{table*}

\subsection{Baselines and Models}
To evaluate the effectiveness of our approach to mine implicit feedback, we set the following two baselines.
\begin{itemize}
    \item \textbf{Original}: Only the human labeled data is used to train the QA model.
    
    \item \textbf{FBQA\textsubscript{ctr}}: The FeedbackQA\textsubscript{ctr} data is used for pre-training the QA model at the first stage. At the second stage, the QA model is further fine-tuned using the human labeled data.
\end{itemize}

In our approach, the best performing feedback model GBDT in Table~\ref{table:metrics for different modeling method} is used to auto-label large scale pre-training data (i.e. FeedbackQA\textsubscript{gbdt}) for pre-training the QA model. At the second stage, the QA model is further fine-tuned using the human labeled data. This approach is referred to as \textbf{FBQA\textsubscript{FA}} in the later experiment results. 

We build the QA relevance models based on two popular deep neural networks, BiLSTM and BERT\textsubscript{base}\footnote{Our goal is to verify the effectiveness of approach, so we do not use BERT\textsubscript{large}, which is time and resource consuming.}. The detailed description of these two models as well as the experimental setting are presented in Appendix~\ref{sec:append_exp_setting}.

\subsection{Results and Discussions}
\label{sec_result}

\subsubsection{Overall Comparison Results}
Table~\ref{table:result} shows the experimental results across all settings. We observe the following.  
\begin{itemize}
    \item Compared with the two baselines Original and FBQA\textsubscript{ctr}, our implicit feedback approach FBQA\textsubscript{FA} achieves significant improvements over different pre-training data size $\{0.5m$, $1m, 4m\}$ and different QA fine-tuning data size $\{5k, 10k, 20k,$ $30k\}$. When the size of feedback pre-training data reaches $4m$, our model gets the best results on the experiment set: for BiLSTM, there is about 5 AUC points increase on average; for BERT, 6 AUC points increase on average.
    
    \item Especially for low resource settings such as $5k$ and $10k$ QA fine-tuning data, our approach shows excellent results to save the labeling cost. Take BERT setting as an example. When the size of pre-training data  equals to $4m$ and fine-tuning data equals to $5k$, our model can get 76.53 AUC metric, which is even higher than the Original result on $30k$ fine-tuning data. In other words, with only $1/6$ of the human labeled data, our model can still outperform the model trained on the full dataset. This experiment verifies the effectiveness of our approach to save large labeling cost. 
    
    \item When we increase the size of implicit feedback pre-training data from $0.5m$ to $1m$ and $4m$, our model is able to get consistent gains in all experimental settings. While for FBQA\textsubscript{ctr}, the gains are not consistent: increasing pre-training data size does not necessarily improve the metrics, which aligns with our findings in Section~\ref{sec:imp_feedback_modeling}. 
    
    \item Our approach shows substantial gains over the baselines with both the BiLSTM and BERT models, which verifies the model agnostic characteristic of our approach. It is expected that BERT based QA models outperform BiLSTM based models since BERT benefits from large scale unsupervised pre-training as well as a large size of model parameters. It is interesting to find that even on top of the powerful deep pre-trained model such as BERT, further significant gains can be obtained. This demonstrates the huge potential of the inexpensive, abundant implicit feedback derived from large-scale user behavior data as complementary data sources to the expensive human labeled data of a relatively small size.
    
\end{itemize}

\subsubsection{Effect of Pre-training Data Size}
\begin{figure}[t]
\setlength{\belowcaptionskip}{-0.1cm}
\centering
\includegraphics[scale=0.36, viewport=2 5 600 410, clip=true]{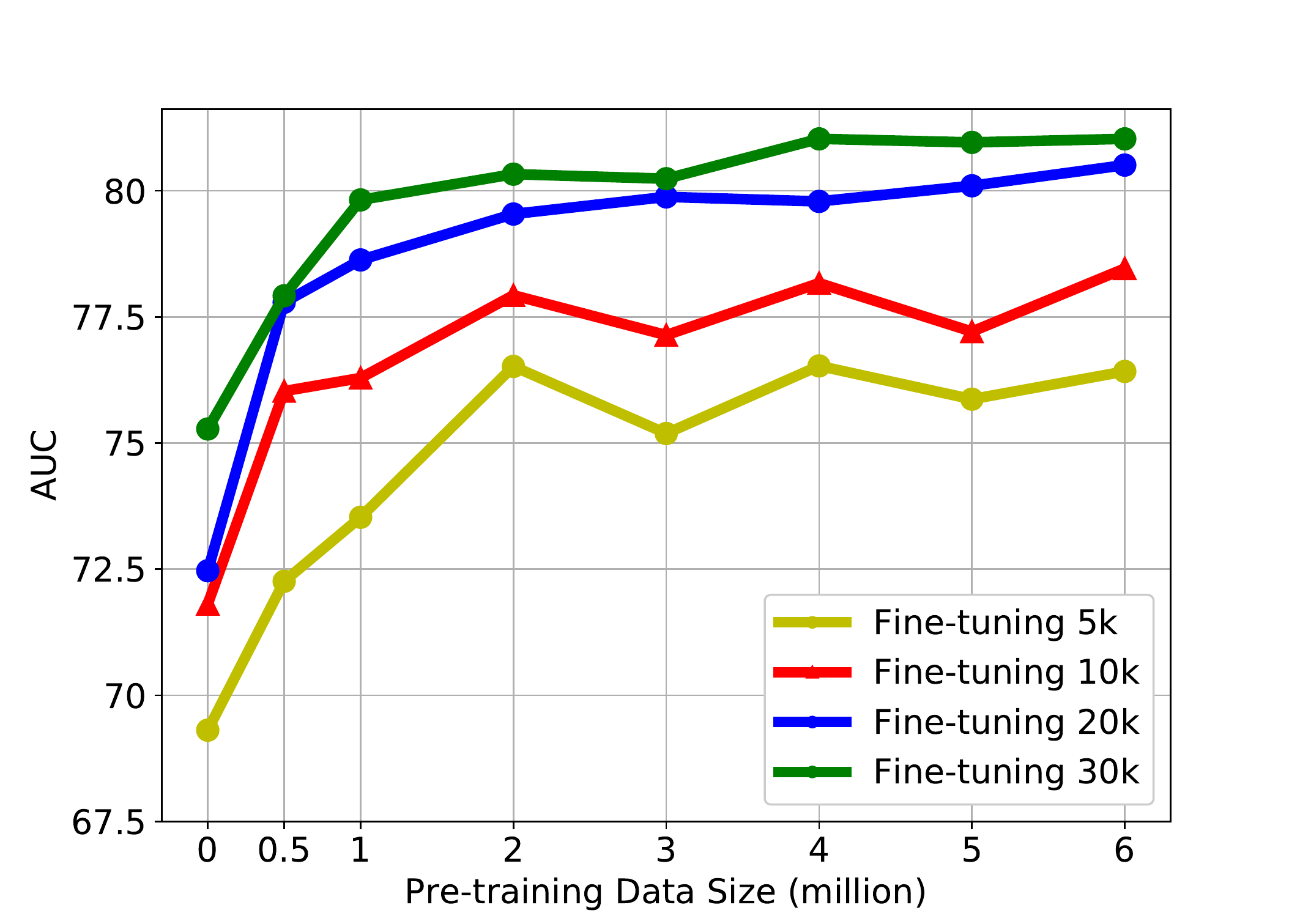}
\caption{\label{figure:feedbackdatasize_auc} Performance on dataset DeepQA  with different FeedbackQA pre-training data size.}
\label{fig:impact_size}
\end{figure}

To further analyze the effect of user implicit feedback on improving web QA, we explore the model performance with respect to the size of the feedback data employed in the pre-training stage. The experiments are conducted on the DeepQA dataset using BERT\textsubscript{base} models. Pre-training data size is set to \{0, 1, 2, 3, 4, 5, 6\} millions. The results are shown in Figure~\ref{figure:feedbackdatasize_auc}. By increasing the size of implicit feedback data in pre-training from 0 to 4 millions, the model performance improves accordingly. However, when the data size reaches a certain scale, e.g, 4 millions in our experiments, the AUC metric on test set slowly flattens out. This suggests that the noise in the implicit feedback data may limit the growth of the improvement.

\subsubsection{Results on the MS Marco and WikiPassageQA datasets}
We further apply our pre-trained model (trained on 4m FeedbackQA\textsubscript{gbdt}) from the implicit relevance feedback data to two open benchmark QA datasets, MS Marco and WikiPassageQA. The results are reported in Table~\ref{table:marco_test}. We find that the queries in the MS Marco dataset are simpler than those in the DeepQA dataset, consequently, with 10k human labeled data and the BERT model, the AUC can reach as high as 94.01\%. Our approach also shows improvement on this dataset, although the gain is not as large as that on the DeepQA dataset. For the BiLSTM model, the gain is around 2 points on average in terms of AUC over 2k, 5k, and 10k human labeled training data in the fine-tuning stage. For the BERT model, since the baseline is already very strong, the improvement is less than 1 point. On the WikiPassageQA dataset, our approach also shows gains, with around 1.3 points AUC gains for BERT and around 1.8 points AUC gains for BiLSTM on average over 2k, 5k and 10k labeled fine-tuning data. 

\begin{table}[t]
\small
\caption{\label{table:marco_test} Comparison Results on MS Marco and WikiPassageQA datasets (all AUC metrics are percentage numbers with \% omitted).}
\begin{tabular}{ccccc|ccc}
\hline
\multirow{2}{*}{\textbf{Model}}  & \multirow{2}{*}{\textbf{Method}}  & \multicolumn{3}{c|}{\textbf{MS Marco}} & \multicolumn{3}{c}{\textbf{WikiPassageQA}} \\
                                 &                                   & \textbf{2k}  & \textbf{5k}  & \textbf{10k}  & \textbf{2k}  & \textbf{5k}  & \textbf{10k} \\ \hline
\multirow{3}{*}{\textbf{BiLSTM}} & \textbf{Original}                                &  64.70             & 64.25              & 65.61   & 55.39  & 58.37  & 60.72          \\
                                 & \textbf{FBQA\textsubscript{ctr}}                 &  62.52             & 65.50              & 66.46   & 53.80  & 58.95  & 61.16         \\
                                 & \textbf{FBQA\textsubscript{FA}}                  &  \textbf{65.65}             & \textbf{66.24}              & \textbf{68.66}    & \textbf{57.23} & \textbf{60.38} & \textbf{62.30}          \\ \hline
\multirow{3}{*}{\textbf{BERT}}   & \textbf{Original}                                &      87.93          &     93.02          &    94.01     & 78.03 & 83.70 & 84.70 \\
                                 & \textbf{FBQA\textsubscript{ctr}}                 &      88.70           &       93.42        &       94.47   & 78.04 & 81.18 & 85.66     \\
                                 & \textbf{FBQA\textsubscript{FA}}                  &   \textbf{88.75}            &   \textbf{94.02}            &     \textbf{94.81}    & \textbf{80.10} & \textbf{84.14} & \textbf{86.16}     \\ \hline
\end{tabular}
\end{table}

\nop{
\subsubsection{Comparison of objective functions in the pre-training stage}
\label{subsec:objective impact}
Once we learn an implicit relevance feedback model, we can apply it to a query passage pair to predict a relevance score. In Equation~\ref{eq:label}, we use two thresholds to convert this continuous relevance score into a Boolean relevance label. In this section, we would like to explore whether it is better to fit the Boolean relevance label in the pre-training stage, or alternatively, fit the original continuous relevance score (between 0 and 1). To fit the Boolean relevance label, we carry out a classification task and use the loss of cross-entropy (CE) as in Equation~\ref{eq:cross_entropy}. Instead, to fit the original continuous relevance score, we conduct a regression task and use the mean squared error (MSE) as the loss function: 
\begin{align}
&L_{MSE}=\frac{1}{n}\sum_{i}^{n}({y}_{i}-\bar{y}_{i})^{2}
\end{align}
where $y_{i}$ represents our QA model output, $\bar{y}_{i}$ represents the auto-labeled relevance score output by the feedback model in Table~\ref{table:metrics for different modeling method}, and $n$ represents the number of training samples.

We compare the results from the two objective functions in Figure~\ref{figure:impact_obj}. It is clear that for both BERT and BiLSTM, the performance of using regression loss in the pre-training stage is consistently worse than that adopting the cross-entropy loss. It indicates that the user feedback data is noisy, and learning the fine-grained data distribution with regression task may propagate the noise to the fine-tuning stage. Instead, the Boolean label approach considers the implicit relevance feedback as a rough indicator, and throws away the uncertain cases falling between the two thresholds. Consequently, the Boolean label approach is more robust to the noise. 

\begin{figure}[H]
\setlength{\belowcaptionskip}{-0.2cm}
\centering
\includegraphics[scale=0.22, viewport=20 20 680 540, clip=true]{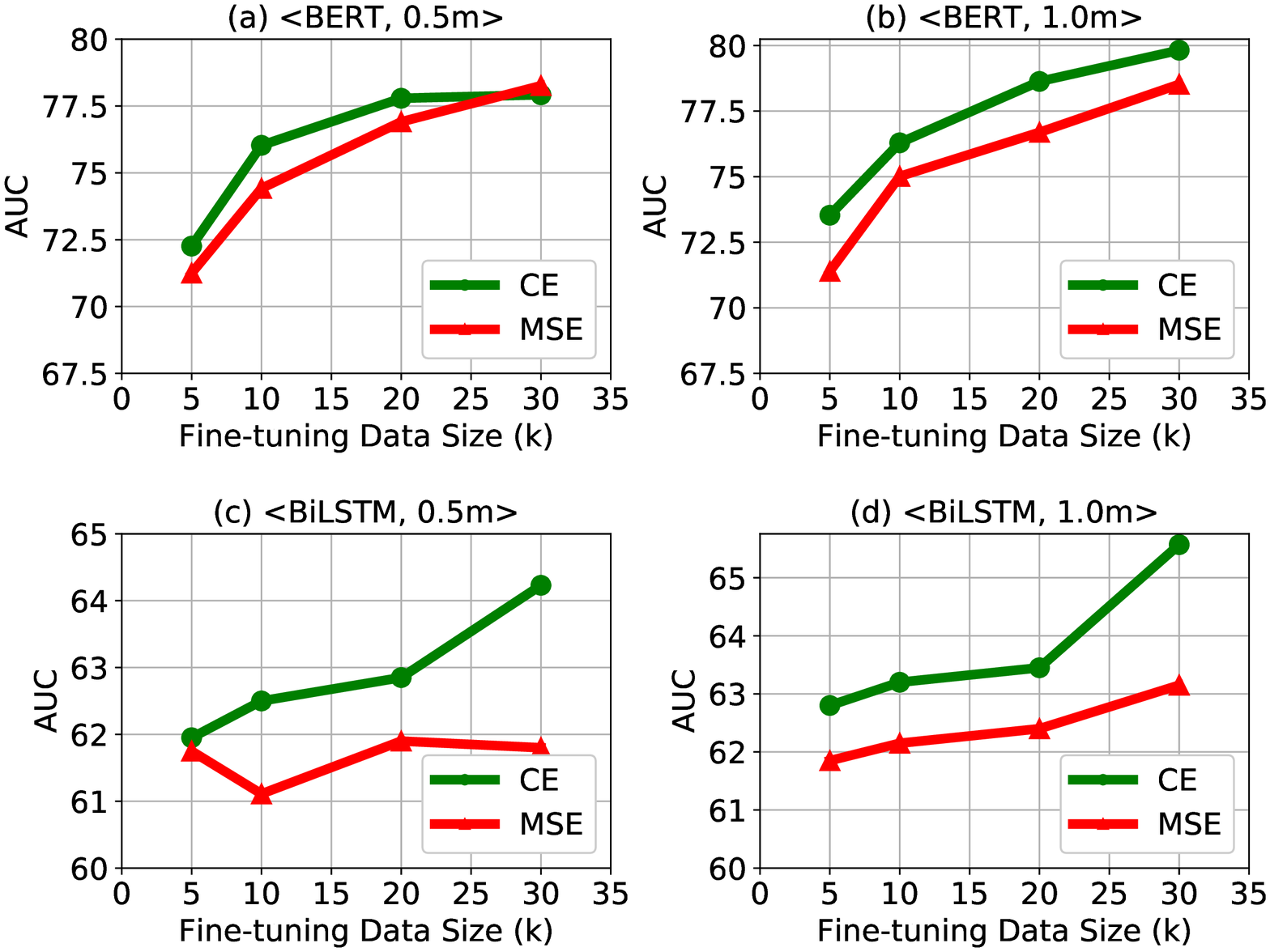}
\caption{\label{figure:impact_obj} Comparison of objective functions during the QA pre-training stage on DeepQA\textsubscript{general} dataset.}
\end{figure}
}

\subsection{Applications to non-English QA}
We further apply our approach to several non-English markets in a commercial search engine. We find user behavior in different countries is very consistent. Consequently, the implicit relevance feedback model trained in en-US market can be successfully transferred to foreign markets without any tuning. The results are shown in Table~\ref{t:frde_qa}. 

In the de-DE (German) and the fr-FR (French) markets, our approach significantly improves the QA service in AUC metric, saving a huge amount of human labeling cost. Take fr-FR as example, our approach shows around 3.2 consistent AUC gains across all training data sizes. Meanwhile, the FBQA\textsubscript{FA} model is able to get 76.43 AUC with only 5k training data while the Original model needs 30k training data to get similar results.   

\begin{table}[t]
    \small
    \caption{\label{t:frde_qa} Results on French and German QnA.}
    \centering
    \begin{tabular}{@{}ccccc@{}}
    \hline
    \textbf{Model} & \multicolumn{4}{c}{\textbf{AUC of fr-FR \& de-DE}}                                       \\
                   & \textbf{5k} & \textbf{10k} & \textbf{30k} & \textbf{50k} \\ \midrule
    \textbf{Original}                     & 73.05/71.46              & 73.99/73.15             & 76.23/75.84       & 76.82/77.11      \\
    
    
    \textbf{FBQA\textsubscript{FA}}  & {\ul \textbf{76.43/76.64}}      & {\ul \textbf{77.26/76.22}}             & {\ul \textbf{79.28/78.83}}   &  {\ul \textbf{80.31/79.76}}      \\ 
    
    \hline
    \end{tabular}
\end{table}

\section{Conclusion and Future work}
\label{sec:conclusion}
This paper proposes a novel framework of mining implicit relevance feedback from user behavior. The implicit feedback models are further applied to generate weak supervised data to train QA models. Our extensive experiments demonstrate the effectiveness of this approach in improving the performance of QA models and thus reducing the human labeling cost. 

Mining implicit feedback from user behavior data for web QA task is an interesting area to explore. In this study, we mainly focus on users' search behavior.  As future work, we may combine users' search behavior with browse behavior. Moreover, we may also conduct deeper analysis on the question types and compare the effectiveness of implicit feedback on different types of queries. Understanding when to trigger QA block from user feedback is another interesting problem. Finally, the application of our approach to more languages is also in our future plan.


\bibliographystyle{ACM-Reference-Format}
\bibliography{sample-base}

\newpage
\appendix
\section{More description of data sets}\label{sec:append_data}
In the following, we present more details about the datasets in Section~\ref{sec:exp_data}. We add one extra dataset called DeepQA\textsubscript{factoid}. For clarity, the original data DeepQA is renamed as DeepQA\textsubscript{general}. The statistics are summarized in Table~\ref{t:append_statistic}.

\begin{itemize}
    \item \textbf{DeepQA\textsubscript{general}}: An English QA dataset where each case consists of three parts, i.e. question, passage, and a binary label (i.e. 0 or 1) by crowd sourcing human judges. The data collection process is as follows. First, for each question $Q$, all the passages $P$ extracted from the top 10 relevant web documents returned by the search engine are collected to form a candidate set of $\langle Q, P\rangle$ pairs. Next, each $\langle Q, P\rangle$ pair is sent to three crowd sourcing judges, and a label (1 for relevance and 0 otherwise) is derived based on a majority voting. 
    
    \item \textbf{DeepQA\textsubscript{factoid}}: This dataset is collected in a similar process as DeepQA\textsubscript{general}. The main difference is that the queries in DeepQA\textsubscript{factoid} are mainly factoid queries, i.e., queries asking about ``what'', ``who'', ``where'', and ``when''. 
    
    \item \textbf{MS Marco}: An open source QA dataset~\cite{DBLP:conf/nips/NguyenRSGTMD16}, which contains questions generated from real anonymized Bing user queries. Each question is associated with multiple passages extracted from the Bing web search results. Well trained judges read the question and its related passages, if there is an answer present, the supporting passages are annotated as relevant, while others are labeled as irrelevant. To evaluate the effectiveness of our approach under low-resource setting (i.e., only a small amount of labeled data are available), the dataset is further sub-sampled to form a positive/negative balanced set with 10k/1k/1k as training, dev and testing sets, respectively. 
    
    \item  \textbf{WikiPassageQA}: An open source QA dataset~\cite{DBLP:conf/sigir/CohenYC18} based on Wikipedia, targeting non-factoid passage retrieval tasks. It contains thousands of questions with annotated answers. Each question is associated with multiple passages in the documents.  In order to obtain a dataset of low resource setting, the dataset is further sub-sampled to form a positive/negative balanced set with 10k/1k/1k as training, dev and testing sets, respectively.
\end{itemize}

\begin{table}[H]
\begin{center}
\caption{\label{t:append_statistic} Statistics of experiment datasets.}
\begin{tabular}{lcccc}
\hline
\textbf{Dataset}          & \textbf{Train} & \textbf{Dev} & \textbf{Test} & \textbf{Labels}\\ 
\hline
\textbf{DeepQA\textsubscript{factoid}}               & 30k   & 2k   & 2k    & 55.7\%+/44.3\%-       \\
\textbf{DeepQA\textsubscript{general}}               & 30k   & 2k   & 2k    & 57.6\%+/42.4\%-          \\
\textbf{MS Marco} &    10k         &      1k      &     1k  & 50\%+/50\%-                    \\ 
\textbf{WikiPassageQA} &    10k         &      1k      &     1k  & 50\%+/50\%- \\
\hline
\end{tabular}
\label{dataset}
\end{center}
\end{table}

\section{More details on experimental setting}\label{sec:append_exp_setting}
We build the QA relevance models based on the following two popular deep neural networks.
\begin{itemize}
    \item \textbf{BiLSTM}. It consists of three parts. The first is an embedding layer, which maps each token to a vector with a fixed dimension. The second is multi-layer Bidirectional LSTM, which encodes both questions and passages based on the token embeddings, i.e. $H^q = BiLSTM_1(Q)$ and $H^p = BiLSTM_2(P)$, where $H^q$ and $H^p$ are the representations for a question and a passage respectively. The parameters of $BiLSTM_1$ and $BiLSTM_2$ are not shared. Following the BiLSTM layer is a prediction layer, which includes a combination layer to concatenate $H^q$ and $H^p$, and then a fully connection layer to predict the relevance of the passage and the question. 
    \item \textbf{BERT\textsubscript{base}}: It contains 12 bidirectional transformer encoders. We concatenate the question text and the passage text together as a single input of the BERT encoder. We then feed the final hidden state of the first token (\emph{[CLS]} token embedding) from the input into a two-layer feed-forward neural network. The final output is the relevance score between the input question and the passage. In all cases, the hidden size is set to 768. The number of self-attention heads is set to 12, and the feed-forward filter size is set to 3072. 
\end{itemize}

For the experiments with BiLSTM, we use two BiLSTM layers with the hidden size of BiLSTM cell set as 128. We set the maximum length of questions and maximum length of passages to be 30 and 200, respectively, for data sets DeepQA, MS Marco, WikiPassageQA and FrenchGermanQA. We set the batch size to 256 and the dimension of word embedding to 300. During pre-training, the learning rate is set to \{1e\textsuperscript{-5}, 3e\textsuperscript{-5}, 5e\textsuperscript{-5}\}, the dropout rate to $\{0.1, 0.3, 0.5\}$ and the max epoch to 50. We choose the best model for fine-tuning based on the evaluation metric on the dev set. During fine-tuning, the learning rate is set to \{1e\textsuperscript{-5}, 3e\textsuperscript{-5}, 5e\textsuperscript{-5}\}, the dropout rate to 0.5 and the max epoch to 50. The best model is selected based on the evaluation metric on the dev set as well.

For the experiments with BERT\textsubscript{base}, we use the huggingface version pre-trained BERT\textsubscript{base} model\footnote{\url{https://github.com/huggingface/pytorch-transformers}}. We set the maximum sequence length to 200 for datasets DeepQA, MS Marco, WikiPassageQA and FrenchGermanQA. We set the batch size to 128, the number of gradient accumulation steps to 2 and the learning rate warmup ratio to 0.1. During pre-training, the learning rate is set to \{1e\textsuperscript{-5}, 3e\textsuperscript{-5}, 5e\textsuperscript{-5}\} and the max epoch to 3. We choose the best model for fine-tuning based on the evaluation metric on the dev set. During fine-tuning, the learning rate is set to \{1e\textsuperscript{-5}, 3e\textsuperscript{-5}, 5e\textsuperscript{-5}\}, and the max epoch to 3. The best model is selected based on the evaluation metric on the dev set as well. 


\begin{table*}[t!]
\small
\caption{Results on DeepQA\textsubscript{factoid}. All ACC and AUC metrics in the table are in percentage , where the sign \% are omitted.}\label{t:deepqa_factoid_results}

\begin{tabular}{ccclllll}
\hline
\multirow{2}{*}{\textbf{Model}}                    & \multirow{2}{*}{\textbf{Method}}                 & \multirow{2}{*}{\textbf{\begin{tabular}[c]{@{}c@{}}Pre-training\\ Data Size\end{tabular}}} & \multicolumn{4}{c}{\textbf{Performance on Different Fine-tuning Data Size (AUC/ACC)}}                                                                  \\
\textbf{}                         & \textbf{}                        & \textbf{}       & \textbf{5k} & \textbf{10k} & \textbf{20k} & \textbf{30k}               \\ \hline

\multirow{7}{*}{\textbf{BiLSTM}}  & \multirow{1}{*}{\textbf{Original}} & \textbf{-}   & 64.02/60.80                         & 64.73/59.00                         & 65.18/60.80                     & 64.03/58.95 \\  \cdashline{3-7}
                                  
                                  & \multirow{3}{*}{\textbf{FBQA\textsubscript{ctr}}} 
                                                                     & \textbf{0.5m}  & 63.62/60.35 (-0.40/-0.45)            & 65.51/59.00 (+0.78/+0.00)            & 64.56/60.40 (-1.24/-0.40)        & 61.79/58.05 (-2.24/-0.90) \\
                                  &                                  & \textbf{1.0m}    & 63.67/60.30 (-0.35/-0.50)            & 64.86/58.65 (+0.15/-0.35)            & 65.53/60.40 (+0.35/-0.40)        & 61.35/57.40 (-2.68/-1.55) \\
                                  &                                  & \textbf{4.0m}    & 62.83/60.20 (-1.19/-0.60)            & 65.09/59.05 (+0.36/+0.05)            & 64.67/59.70 (-0.51/-1.10)        & 65.24/60.95 (+1.21/+2.00) \\ \cdashline{3-7}
                                  
                                  & \multirow{3}{*}{\textbf{FBQA\textsubscript{FA}}} 
                                                                     & \textbf{0.5m}  & 65.66/62.05 (+1.64/+1.25)            & 67.87/63.00 (+3.14/+4.00)            & 68.84/63.80 (+3.66/+3.00)        & 69.80/64.75 (+5.77/+5.80) \\
                                  &                                  & \textbf{1.0m}    & 66.36/64.00 (+2.34/+3.2)             & 67.81/63.45 (+3.08/+4.45)            & 69.23/64.70 (+4.05/+3.90)        & 72.16/66.40 (+8.13/+7.45) \\
                                  &                                  & \textbf{4.0m}    & \textbf{67.22/65.25 (+3.20/+4.45)}   & \textbf{70.26/65.10 (+5.53/+6.10)}   & \textbf{70.79/64.95 (+5.61/+4.15)}        
                                  & \textbf{73.31/67.15 (+9.28/+8.20)} \\
                                  
\hline 

\multirow{7}{*}{\textbf{BERT}}    & \multirow{1}{*}{\textbf{Original}} & \textbf{-}   & 68.88/65.46               & 71.43/66.77                 & 73.87/67.06                     & 78.37/69.56 \\  \cdashline{3-7}

                                  & \multirow{3}{*}{\textbf{FBQA\textsubscript{ctr}}} 
                                                                     & \textbf{0.5m}  & 71.44/66.27 (+2.56/+0.81)  & 75.17/67.06 (+3.74/+0.29)    & 77.78/70.78 (+3.91/+3.72)        & 80.00/72.87 (+1.63/+3.31) \\
                                  &                                  & \textbf{1.0m}    & 71.73/66.17 (+2.85/+0.71)  & 74.19/67.26 (+2.76/+0.49)    & 77.01/69.46 (+3.14/+2.40)        & 78.30/72.07 (-0.07/+2.51) \\
                                  &                                  & \textbf{4.0m}    & 70.55/66.06 (+1.67/+0.60)  & 75.51/68.26 (+4.08/+1.49)    & 78.10/69.77 (+4.23/+2.71)        & 79.54/70.97 (+1.17/+1.41) \\ \cdashline{3-7}
                                  
                                  & \multirow{3}{*}{\textbf{FBQA\textsubscript{FA}}} 
                                                                     & \textbf{0.5m}  & 74.00/67.77 (+5.12/+2.31)  & 76.57/68.87 (+5.14/+2.10)    & 78.78/70.67 (+4.91/+3.61)       & 80.91/72.37 (+2.54/+2.81) \\
                                  
                                  &                                  & \textbf{1.0m}    & 74.33/68.67 (+5.45/+3.21)  & 75.77/68.77 (+4.34/+2.00)    & 78.40/71.17 (+4.53/+4.11)       & 79.31/71.77 (+0.94/+2.21)  \\
                                  
                                  &                                  & \textbf{4.0m}    & \textbf{76.13/69.77 (+7.25/+4.31)}   & \textbf{77.93/71.27 (+6.50/+4.50)}   & \textbf{79.94/72.77 (+6.07/+5.71)}        & \textbf{81.21/73.27 (+2.84/+3.71)} \\

              \hline
\end{tabular}
\end{table*}

\section{More experimental results and discussions}
We conduct further experiments on DeepQA\textsubscript{factoid} dataset. The results are shown in Table \ref{t:deepqa_factoid_results}. Similar to DeepQA\textsubscript{general} dataset, our implicit feedback approach FBQA\textsubscript{FA} achieves significant improvements over different pre-training data size $\{0.5m$, $1m, 4m\}$ and different QA fine-tuning data size $\{5k, 10k, 20k,$ $30k\}$. For comparison, click based FBQA\textsubscript{ctr} model is not able to get consistent gains compared with Original baseline, which proves that click does not necessarily suggest relevance in terms of QA task. 

The results further demonstrate the effectiveness of our proposed implicit feedback QA approach by leveraging large-scale user behavior data as complementary data sources to save the human labeling cost.

\end{document}